\documentclass[10pt,a4paper,onecolumn]{article}
\usepackage{graphicx}
\usepackage{bm}
\usepackage{color}
\date{}

\begin{document}

\title{Attainability of maximum work and the reversible efficiency from 
minimally nonlinear irreversible heat engines }

\author{M. Ponmurugan \\
Department of Physics, School of Basic and Applied Sciences, \\
Central University of Tamilnadu, Thiruvarur - 610 005,  \\
Tamilnadu, India. e-mail:ponphy@cutn.ac.in}


\maketitle

\begin{abstract}

We use the general formulation of irreversible thermodynamics 
and study the minimally nonlinear irreversible model of heat engines operating between a time-varying   
hot heat source of finite size and a cold heat reservoir of infinite size.  
We  find  the criterion in which the optimized efficiency obtained by this minimally nonlinear irreversible 
heat engine can reach the reversible efficiency under the tight coupling condition: a condition of no
heat leakage between the system and the reservoirs. 
We assume the rate of heat transfer from hot to cold heat reservoir obeys Fourier law and 
discuss physical conditions under which one can obtain the reversible 
efficiency in a finite time with finite power. 
We also calculate the efficiency at maximum power from the minimally nonlinear irreversible  heat engine 
under the non-tight coupling condition.

\end{abstract}

\section{Introduction}

The theory of irreversible thermodynamics \cite{oldpap,recentpap,prig,neqbook,seifprx} 
nowadays attracts more interest towards 
the formulation of a new theoretical framework as well as the experimental study of the biological 
systems and bio-inspired artificial nanosystems \cite{how,bath,maxpoweruniloss}. 
Most of these systems are highly nonlinear and works under the general 
principle of a heat engine operating in  nonequilibrium conditions.
A heat engine is  a thermodynamic system operating between  two heat reservoirs which consumes heat $Q_h$
from the hot heat source at a given temperature $T_h$ and converts
 part of it as useful work $W$ 
and the remaining heat $Q_c$ is delivered to the cold heat reservoir at a given temperature $T_c$.

Traditional studies of heat engine are based on the reversible thermodynamics formulation of a linear system 
operating between the hot and cold reservoirs of infinite size.  
For an irreversible thermodynamics, most of the studies on heat engine are formulated for the linear system 
operating between the hot and cold heat source of infinite size \cite{vanonsager,carnotfinite}. 
These studies mainly focus on obtaining the efficiency $\eta=\frac{W}{Q_h}$, at maximum power 
in a finite time and its universality behavior
$\eta_U \equiv \eta_C/2+ \eta^2_C/8+O(\eta_C^3)$, \cite{TUuniv,vanmaxpower1,seifmaxpower,vanmaxpower2}
where $\eta_C=1-T_c/T_h$ is the Carnot efficiency of the reversible heat engine with 
zero power. 
The Carnot efficiency, also called the reversible efficiency,
is the maximum efficiency obtained in a quasi-static process, which takes an infinite time for completion. 

For an optimized thermal engine in the endoreversible limit, the efficiency at maximum power 
 given by  $\eta_{CA}=1-\sqrt{1-\eta_C}$ \cite{novikov,curzon} is usually called as the 
Curzon-Ahlborn efficiency.  When the temperature difference between the two reservoirs is small, 
the Taylor expansion of $\eta_{CA}$ gives $\eta_U$ \cite{uniTU,endoqm} which
is bounded below the Carnot efficiency of the reversible heat engines.
It has been shown that the efficiency at maximum power does not show universality behavior even in the 
linear response region of  certain systems \cite{maxpoweruniloss}. The higher values of efficiency obtained by the practical heat engine are not necessarily in the region of maximum power output \cite{nonmaxpow}. Further, a recent study showed that the universal bounds on efficiency can also be derived for an arbitrary 
power \cite{nonmaxpowstdy}.

The general theory of linear irreversible heat engines working between a finite sized hot heat source
and an infinite sized cold reservoir has been formulated recently \cite{bathfinite}. 
This formulation was based on the extraction of maximum  work  called  Exergy \cite{callen} obtained from 
the finite sized hot heat source of time dependent temperature $T$ until the system reaches the final 
equilibrium state of cold reservoir. More general formulation of optimized maximum work output 
and the universal feature of the efficiency at maximum power for the irreversible heat engines
operated  between  finite sized 
heat reservoirs beyond linear regime have been studied very recently \cite{nonlinbath}.

Various studies on heat engines favored the attainability of Carnot efficiency at nonzero power 
\cite{fav,infreservoir1,infreservoir2}, however, it has been ruled out for large classes of systems 
which are in the linear response regime \cite{seifprx,vanreversible,unfav}. This may raise the question of whether it can be reachable for a nonlinear irreversible system at finite power \cite{nonlinbath}. In order to answer this question we have taken the minimally nonlinear irreversible thermodynamic model \cite{nonlinheat, nonlinrefrig,nonlincombi1,nonlincombi2} in our study. 
Although the minimally nonlinear model has been studied partially in Ref. \cite{nonlinbath} by using perturbation method and also for infinite reservoirs \cite{infreservoir1,infreservoir2}, we use this model in exergy study and explicitly calculate the condition to obtain the optimized work and the maximum efficiency. We find that the condition derived resembles with the Eq.(26) of Ref. \cite{nonlinbath}.  

This paper has been organized as follows. In section 2, we introduce the minimally nonlinear 
irreversible model for exergy calculation. In section 3, we incorporate thermodynamical 
optimization procedure and calculate the optimized efficiency under the  tight coupling condition: a condition  
in which there is no heat leakage between the system and the reservoirs \cite{bathfinite,kedem}. 
We also calculate the efficiency at maximum power under the non-tight coupling condition in section 4
and finally, conclude with the main results.

\section{Minimally nonlinear irreversible model}

Incorporating the Onsager relation in the study of heat engines \cite{vanonsager,onsager}, a minimal model 
for a nonlinear heat engine has been introduced by Izumida and Okuda \cite{nonlinheat,nonlinrefrig} which is 
given by
\begin{eqnarray}\label{extons1}
J_1&=&L_{11}X_1+L_{12}X_2 
\end{eqnarray}
\begin{eqnarray}\label{extons2}
J_2&=&L_{21}X_1+L_{22}X_2-r_h J_1^2,
\end{eqnarray}
where $J_i$ is thermodynamic flux and $X_i$ is its conjugate thermodynamic
force which is defined as
\begin{eqnarray}
J_1 &\equiv&  \dot{x}, \\
X_1 &\equiv& F/T_c, \\
J_2 &\equiv& \dot{Q_h}, \\
X_2 &\equiv&  \frac{1}{T_c}-\frac{1}{T},
\end{eqnarray}
and $L_{ij}'s$ are the Onsager coefficients 
with the reciprocity relation $L_{12}=L_{21}$ \cite{bathfinite}. 
For the non-negativity of the entropy production rate \cite{bathfinite,nonlinheat}, 
the possible values of $L_{ij}$ are restricted as $L_{11} \ge 0$, $L_{22} \ge 0$ and 
$L_{11}L_{22}-L_{12}L_{21} \ge 0$.  
The non linear term $r_h J_1^2$  was introduced  to account for the dissipation effect with $r_h>0$ in the 
Onsager relation \cite{vanonsager,nonlinheat}. In the above equations, $F$ denotes  
the time independent external generalized force and $\dot{x}$ denotes 
the time derivative of its conjugate variable $x$. The terms $\dot{Q_h}$, 
$\dot{Q_c}$ and $\dot{W}$ are the time derivatives of $Q_h$,$Q_c$ and $W$ respectively.
The flux $J_1$ related to the time derivative of work as 
$\dot{W}=-F\dot{x}=-X_1T_cJ_1$ \cite{bathfinite}.

Although the dissipation effects due to friction on the heat devices have not been taken
into account in the minimally nonlinear irreversible model, a recent study obtained  a 
clear interpretation of the global performance of generic heat devices using
this model \cite{nonlincombi1}. For more detail studies on this model 
see, Refs.\cite{nonlinheat,nonlinrefrig,nonlincombi1,nonlincombi2}

In order to find out the  optimized efficiency, we use the extended Onsager 
relation as described  in Eqs.(\ref{extons1}) and (\ref{extons2}) 
and study the exergy of nonlinear irreversible heat engines operating 
between a time-varying  hot heat source of finite size and a 
cold heat reservoir of infinite size. The system finally reaches the thermal equilibrium state 
with a uniform temperature of the cold heat reservoir. We calculate the condition for obtaining 
the optimized efficiency as follows.

In our study, we consider the size of the hot reservoir is finite and its temperature evolves from $T_h$ to $T_c$ in a
time interval $\tau$. The time-varying hot heat source which is initially in equilibrium at temperature $T_h$
is assumed to be always in equilibrium for any values of temperature $T$ at the later time. 
The heat capacity at the constant volume 
at any temperature is $C_v=C_v(T)$ and the initial internal energy  and entropy  are $U_h$ and $S_h$ respectively. 
When the hot heat source approaches the final temperature of the cold  reservoir in a time interval 
$0$ to $\tau$, one can calculate the total work extracted by the heat engine as
$W=\int dW =\int \eta^T dQ_h$ where $dQ_h$ is the infinitesimal heat that can be transformed 
into the infinitesimal work $dW$ with the efficiency $\eta^T$ at each $T$. 
This work can be bounded by the Carnot efficiency $\eta_C^T=1-T_c/T$ at each $T$ which is 
given by \cite{bathfinite} 
\begin{eqnarray}\label{wbound}
W \le \int \eta_C^T dQ_h &=&-\int_{T_h}^{T_c} \eta_C^T C_v dT \\ \nonumber
&=&(U_h-U_c) -T_c(S_h-S_c) \\ \nonumber 
&\equiv&  E,
\end{eqnarray}
where
\begin{eqnarray}\label{delU}
U_h - U_c \equiv \int_{T_c}^{T_h} C_v dT,   
\end{eqnarray}
\begin{eqnarray}\label{delS}
S_h-S_c \equiv \int_{T_c}^{T_h} \frac{C_v}{T} dT, 
\end{eqnarray} 
$U_c$ and $S_c$ are respectively the internal  energy and entropy  of the 
final equilibrium state of the hot heat source and $E$ is the maximum work 
called as the exergy. The corresponding efficiency $\eta=W/Q_h=W/(U_h-U_c)$ 
is bounded below the maximum value  as \cite{bathfinite} 
\begin{eqnarray}\label{etabound}
\eta \le \frac{E}{U_h-U_c} &=& 1- \frac{T_c(S_h-S_c)}{U_h-U_c}  \\ \nonumber
&\equiv&  \eta_{max},
\end{eqnarray}
where $\eta_{max}$ is the maximum efficiency attained by the engine. We call $\eta_{max}$
as the reversible efficiency which can be obtained naturally 
for any reversible heat engines operating quasi statically 
taking an infinite time to complete the process.

Let $J_3$ denotes the heat flux of the cold reservoir which 
is given by \cite{bathfinite,nonlinheat}
$J_3 \equiv \dot{Q_c}=\dot{Q_h}-\dot{W}=J_2+J_1X_1T_c$.
Using Eq.(\ref{extons1}) one can obtain $X_1=(J_1-L_{12}X_2)/L_{11}$,
then Eq.(\ref{extons2}) and  $J_3$ can be  rewritten as 
\begin{eqnarray}\label{j2f}
J_2=\frac{L_{21}}{L_{11}}J_1 + L_{22}(1-q^2)X_2 - r_hJ_1^2,
\end{eqnarray}
\begin{eqnarray}\label{j3f}
J_3=\frac{L_{21}T_c}{L_{11}T}J_1 + L_{22}(1-q^2)X_2 + r_c J_1^2,
\end{eqnarray}
where $r_c=\frac{T_c}{L_{11}}-r_h$ and $q=\frac{L_{21}}{\sqrt{L_{11}L_{22}}}$  with $|q| \le 1$ 
is the coefficient of the coupling strength \cite{kedem}. 
Under the condition $|q|=1$ called as the tight coupling condition, the second term
$L_{22}(1-q^2)X_2$ known as the heat leakage from the hot heat source to the 
cold heat reservoir vanishes \cite{bathfinite,nonlinheat}.

By using the above relations, the entropy production rate \cite{oldpap,recentpap},
$\dot{S} = -\frac{J_2}{T}+\frac{J_3}{T_c}$ \cite{bathfinite,nonlinheat,nonlincombi1}
can be written as,
\begin{eqnarray}\label{eprod}
\dot{S}=L_{22}(1-q^2)X_2^2+ \left \{ \frac{r_h}{T}+\frac{r_c}{T_c} \right \}J_1^2 \ge 0.
\end{eqnarray}
Since $r_h > 0$ and also the first term in the above equation is greater 
than or equal to zero, one can naturally make an assumption that 
$r_c > 0$ \cite{nonlinheat,nonlincombi1} such that which should ensure 
the non-negativity of  the entropy production.  
In our study, we did not make such an assumption that  
the value of $r_c$ should be greater than zero.
However, in order to make the positive entropy production rate, 
we impose the condition 
\begin{eqnarray}\label{Jouleheatcond}
\left \{ \frac{r_h}{T}+\frac{r_c}{T_c} \right \} \ge 0.
\end{eqnarray}
This condition can be useful for making the correspondence between the minimally nonlinear 
heat engine model and the thermoelectric heat devices with zero magnetic field
(see Eqs.(31 \& 36) of Ref.\cite{OkuCritise}). In such a case the first and second term 
in Eq.(\ref{eprod}) can be  linked  respectively with the heat bypass and the Joule heating 
which are always positive \cite{nonlinheat,OkuCritise,minthermoelect}.
Since $r_c=\frac{T_c}{L_{11}}-r_h$, the above  condition becomes,
 $\frac{1}{L_{11}}-r_h \left \{ \frac{1}{T_c}-\frac{1}{T} \right \}\ge 0$.  
For time-varying hot heat source, the above condition can be rewritten as
\begin{eqnarray}\label{discond}
X_2 L_{11} r_h \le 1.
\end{eqnarray}
Under this condition, the entropy production rate becomes 
zero when $X_2 L_{11} r_h = 1$ and $|q|=1$.

The rate of decrease of temperature $T$ of the hot heat source when 
the heat engine operates from the initial temperature $T_h$ to the  
final temperature $T_c$ is given by \cite{bathfinite,nonlinbath}
\begin{eqnarray}\label{j2cv}
J_2=-C_v\dot{T}.
\end{eqnarray}
The above equation also provides the relation that connects the 
temperature $T$ and time $t$ with $\dot{T} =\frac{dT}{dt} \ne 0$ in general.
Then Eq.(\ref{j2f}) can be written as 
\begin{eqnarray}\label{j2sub}
r_h J_1^2 -  \frac{L_{21}}{L_{11}}J_1- L_{22}(1-q^2)X_2 - C_v\dot{T}=0.
\end{eqnarray}
The above equation can be written simply as
\begin{eqnarray}\label{j2sub1}
r_h J_1^2 - a_0 J_1-g-C_v\dot{T}=0.
\end{eqnarray}
where  $g=L_{22}(1-q^2)X_2$ is the heat leakage term and 
$a_0=\frac{L_{21}}{L_{11}}$. 
In terms of $g$ and $a_0$, Eq.(\ref{j3f}) can be written as 
\begin{eqnarray}
J_3=\frac{T_c}{T}a_0J_1 + g + \left(\frac{T_c}{L_{11}}- r_h \right)J_1^2.
\end{eqnarray}
Using Eq.(\ref{j2sub1}) in the above equation for $J_1^2$ and  
after simplification one can get,
\begin{eqnarray}\label{j3fsub1}
J_3=a_0(\beta -X_2T_c)J_1+\beta g+(\beta -1)C_v\dot{T},
\end{eqnarray} 
where $\beta =\frac{T_c}{L_{11}r_h} =\frac{r_c}{r_h} +1$ \cite{nonlincombi2}.
Using Eq.(\ref{discond}), $\beta =\frac{X_2T_c}{X_2L_{11}r_h}$ can takes value $\ge X_2T_c$ 
and equality holds when $X_2L_{11}r_h=1$.

Since Eq.(\ref{j2sub1}) is quadratic in $J_1$, it has two roots $J_1^+$ and $J_1^-$ which 
are given by  
\begin{eqnarray}\label{j1root}
J_1^{\pm}&=&\frac{a_0}{2r_h}\Bigg[ 1 \pm  \sqrt{1+\frac{4r_h}{a_0^2}(g+C_v\dot{T})} \Bigg]. \\
         &=&\frac{1}{a_0 a_1} \Bigg[ 1 \pm  \sqrt{1+2a_1 (g+C_v\dot{T})} \Bigg].
\end{eqnarray}
where $a_1=2r_h/a_0^2$. 

We consider only the physically acceptable solution 
of $J_1^{+}$ and discarded the other solution $J_1^{-}$, since $J_1 \ne 0$ as $J_2=0$ \cite{nonlinbath}.
Using $J_1=J_1^{+}$, Eq.({\ref{j3fsub1}}) 
can be expressed as a function of $T$ and $\dot{T}$ as
$J_3(T,\dot{T})=a_0(\beta -X_2T_c)J_1^{+} + \beta g+(\beta -1)C_v\dot{T}$.
The above  equation can be rewritten as  
\begin{eqnarray}\label{j3fsub2}
J_3(T,\dot{T})=k \Big[1 + \sqrt{p} \Big]+\beta g+(\beta -1) C_v\dot{T}.
\end{eqnarray}

Here, we have taken $k=\frac{(\beta  -X_2 T_c)}{a_1}$ 
and $p=1+2a_1(g+C_v\dot{T})$ for notational convenience.
In our further calculation, we assume that $\beta $,$a_0$,$a_1$,$C_v$ and the Onsager coefficients 
depend only on the temperature. Therefore, the leakage term $g$ and $k$ 
depends only on $T$, but $p$ depends on both $T$ and $\dot{T}$. Thus, 
$k(T) =\frac{\beta (T) - X_2(T) T_c}{a_1(T)}$ and 
$p(T,\dot{T})=1+2a_1(T) \big( g(T)+ C_v (T)\dot{T} \big)$.

\section{Thermodynamic Optimization}

The heat $Q_h$ and the work output $W$ obtained from $J_2$ and $J_3$ 
in the time interval $0$ to $\tau$ is \cite{bathfinite}
\begin{eqnarray}\label{qhval}
Q_h=\int_0^{\tau} J_2(t) dt = -\int_{T_h}^{T_c} C_v dT = U_h-U_c, 
\end{eqnarray}
\begin{eqnarray}\label{wout}
W=\int_0^{\tau} \dot{W}(t) dt = U_h-U_c  - \int_{0}^{\tau} J_3(t) dt, 
\end{eqnarray}
where $\dot{W}=\frac{dW}{dt}=J_2 - J_3$. Hence, the total power $P$ and the efficiency $\eta$ 
can be obtained as \cite{bathfinite}
\begin{eqnarray}\label{powval}
P=\frac{W}{\tau}=\frac{U_h-U_c  - \int_{0}^{\tau} J_3(t) dt}{\tau}, 
\end{eqnarray}
\begin{eqnarray}\label{etaval}
\eta=\frac{W}{Q_h}=1- \frac{\int_{0}^{\tau} J_3(t) dt}{U_h-U_c}. 
\end{eqnarray}

In order to maximize the work and hence obtain the maximum efficiency,
we express $J_3(t)$ as a function of $T$ and $\dot{T}$ as in Eq.(\ref{j3fsub2}) and then minimize 
the integral  $\int_0^{\tau} J_3(T,\dot{T}) dt$ in the above equation by solving 
the following Euler-Lagrange equation for $T(t)$  \cite{bathfinite,nonlinbath}
\begin{eqnarray}\label{ELeq}
\frac{d}{dt}\bigg(\frac{\partial J_3(T,\dot{T})}{\partial \dot{T}}\bigg)-\frac{\partial J_3(T,\dot{T})}{\partial{T}}=0.  
\end{eqnarray}
After solving the above equation, we  have obtained  the optimization condition as (see, appendix),
\begin{eqnarray}\label{ELfin1}
\frac{d}{dt}\bigg( \frac{\partial Y} {\partial \dot{T}} \bigg) - \frac{\partial Y}{\partial T} =0, 
\end{eqnarray}
where $Y(T,\dot{T})=k[1 + \sqrt{p}]+\beta g$.

Multiplying Eq.(\ref{ELfin1}) throughout by $\dot{T}$, we obtain
\begin{eqnarray}\label{ELfinal}
\frac{d}{dt}\bigg( \dot{T} \frac{\partial Y} {\partial \dot{T}} - Y \bigg) =0. 
\end{eqnarray}
After integrating the above equation one can get
\begin{eqnarray}\label{ELconst}
\dot{T} \frac{\partial Y(T,\dot{T})} {\partial \dot{T}} - Y(T,\dot{T}) = A, 
\end{eqnarray}
where $A$ is a $\tau$ dependent integration constant. 
As similar to Eq.(26) of Ref.\cite{nonlinbath}, 
for any coupling strength $|q| \le 1$, we have obtained the
necessary condition to achieve an optimized work output 
from the  minimally nonlinear irreversible model of heat engines.

It should be noted that the condition in terms of $Y(T,\dot{T})$ obtained 
in our study is  not for the entropy production rate as given in Ref.\cite{nonlinbath}.
Eq.(\ref{ELconst}) is a highly nonlinear implicit differential equation \cite{nondiff}, 
it may be difficult to simplify this equation for further analysis and hence we do not  try  
any other optimization \cite{twoopti} to minimize the integral $\int_{0}^{\tau} J_3(t) dt$ further.
Under this optimization condition Eq.(\ref{j3fsub2}) becomes 
\begin{eqnarray*}\label{j3min0}
J_3(T,\dot{T})&=& \dot{T} \frac{\partial Y(T,\dot{T})} {\partial \dot{T}} - A + (\beta -1) C_v\dot{T} \\
&=& \frac{k}{\sqrt{p}} a_1C_v\dot{T}-A+(\beta -1)C_v\dot{T}.  
\end{eqnarray*}

The condition for positive entropy production rate 
(Eq.\ref{discond}) can be written in terms of $\beta$ as
$X_2 L_{11} r_h = \frac{X_2T_c}{\beta} \le 1$, then
\begin{eqnarray}\label{Betacond}
\beta \ge X_2T_c.
\end{eqnarray}
For the lowest value of $\beta=X_2T_c$, $k=\frac{(\beta -X_2T_c)}{a_1}=0$
and hence from Eq.(\ref{ELconst}) with 
$Y(T,\dot{T})=k[1 + \sqrt{p}] + \beta g$, we get
\begin{eqnarray}\label{Aval}
A &=&   \frac{k}{\sqrt{p}} a_1C_v\dot{T} - \bigg(k[1 + \sqrt{p}] + \beta g \bigg) \\ \nonumber
&=&-X_2 T_c g. 
\end{eqnarray}

Then, the  optimized flux  is given by 
\begin{eqnarray}\label{j3min1}
J_3(T,\dot{T}) &=&  X_2 T_c g + (X_2T_c-1) C_v \dot{T}. 
\end{eqnarray}
The above equation has been obtained  by optimizing $J_3$  with $\beta=X_2T_c$ 
for any value of the coupling strength $|q| \le 1$. For this minimum value of $\beta$,  
the entropy production should be  independent of 
time under the tight coupling condition, $|q|=1$.
In this condition the leakage term $g=0$ and the equation (\ref{j3min1}) becomes
\begin{eqnarray}\label{j3final}
J_3(T,\dot{T})=(X_2T_c-1) C_v\dot{T} = -\frac{T_c}{T}C_v\dot{T}.
\end{eqnarray} 
Integrating the above equation from $0$ to $\tau$ and using Eq.(\ref{delS}), we get
$\int_0^{\tau} {J_3(T,\dot{T})} dt=-T_c\int_{T_h}^{T_c} {\frac{C_v}{T} dT} =T_c (S_h-S_c)$.
By using Eq.(\ref{etaval}), the  optimized efficiency $\eta$ can be obtained as   
\begin{eqnarray}\label{eff1}
\eta&=&1- \frac{\int_0^{\tau} {J_3(T,\dot{T})} dt}{U_h-U_c} \\ \nonumber
&=& 1- \frac{T_c (S_h-S_c)}{U_h-U_c}  \\ \nonumber
&=&\eta_{max}.
\end{eqnarray}
Thus, the reversible efficiency has been obtained from the minimally nonlinear irreversible heat engine 
under the tight coupling condition.
In the case of $C_v \to \infty $,  for an isothermal environment, $\eta_{max}$ recovers the usual Carnot efficiency $\eta_C$  
by the definition \cite{bathfinite} $\frac{U_h-U_c}{T_h}=\frac{Q_h}{T_h}=S_h-S_c$.
The maximum work (Exergy) extracted and also the total power obtained from this nonlinear 
irreversible heat engine are obtained as  
\begin{eqnarray}\label{powvalfin}
W &=&U_h-U_c  - T_c (S_h-S_c) \equiv E, \\ 
P&=&\frac{W}{\tau}=\frac{E}{\tau}. 
\end{eqnarray}

\section {Discussion}
In the above analysis, we found the criterion in which the optimized efficiency obtained by the minimally nonlinear irreversible heat engine can reach the reversible efficiency under the tight coupling condition. In order to discuss physical circumstances under which the value of $\tau$ is finite, one should assume a specific relation between heat current and local temperature gradient. 
Since Fourier's law holds even for the nonlinear regime \cite{fourier}, 
we assume that the rate of heat transfer $dQ(t)/dt$ from hot to cold bath obeys Fourier law.

In the present case, $dQ(t)/dt = \kappa (T(t) - T_c)$, where $\kappa$ is the thermal conductance of 
the material connecting hot and cold bath. The temperature of the hot bath $T(t)$ at time $t$
is connected with $dQ/dt$ by heat capacity $C_v$ as $dQ(t)/dt = - C_v dT(t)/dt$.
Then, the time rate of change of temperature of the hot bath is given by  
$dT(t)/dt = - \gamma (T(t) - T_c)$, where the decay rate  $\gamma = \kappa/C_v$. 
The solution of this equation with the initial condition $T(0) = T_h$ is 
obtained as 
\begin{eqnarray}\label{Fpara}
T(t) = T_c + (T_h - T_c)e^{-\gamma t}.
\end{eqnarray}
This shows that, for an exponential relaxation from $T_h \to T_c$, the equilibration occurs in a finite time only if the relaxation rate $\gamma$ diverges. For a finite $\gamma$, the relaxation time $\tau$ can be assumed to be given by $t$ for which $T$ almost relaxed to $T_c$.  For example, we can take $\tau = 1/\gamma$, a finite time at which the temperature $T(t) = T_c + (T_h - T_c)/e$.  A better way to define $\tau$ is via the formula $T(\tau) = T_c + \epsilon$, where $\epsilon$ would be a given small number. 
Then, the value of $\tau$ in which one can obtain the maximum work and the reversible efficiency is given by 
\begin{eqnarray}\label{Fparatau}
\tau = \frac{1}{\gamma} \ ln\left(\frac{T_h-T_c}{\epsilon}\right).
\end{eqnarray}
It should be noted that the thermal conductance $\kappa$ of a material generally varies with temperature. However, in our discussion, the decay rate $\gamma$ does not vary appreciably over a significant range of temperatures and thus 
$\kappa$ can be treated as a constant.

Our result showed that the reversible efficiency 
obtained from the nonlinear irreversible heat engines under the  tight coupling condition
is not necessarily to be in the regime of maximum or zero power output in a time interval 0 to $\tau$.
Based on the non-zero entropy production rate, it was recently proved that Carnot efficiency at finite 
power is impossible for (a  Markov process description of) a general thermodynamic system even in the nonlinear regime \cite{Marko}. However, in our work, we have achieved the reversible (Carnot) efficiency at finite power as a special case of zero entropy production rate. Although our result is  entirely  based on the positive entropy production  rate condition, $X_2L_{11}r_h \le 1$ (Eq.\ref{discond}) or $\beta \le X_2T_c$ (Eq.\ref{Betacond}), when tight coupling condition is considered, zero entropy production rate is determined by Eq.(\ref{discond}) holding an 
equality  $X_2L_{11}r_h = 1$  ($\beta = X_2T_c$). This equation relates the dissipation constant $r_h$ with the rate
of decrease of the temperature of the hot finite reservoir, which in turn is related with the size of the reservoir and with the time scale of the heat exchange process.

The Onsager symmetry used in our analysis
 reduces the generality of the present result and it is valid only 
for the steady state heat engines. 
 However, for cyclic heat engines, this simplification is permitted only under 
the condition that the driving protocols are
 symmetric under time-reversal \cite{seifprx}.	
Our result (Eq. \ref{eff1}) showed that if we design a practical heat engine whose positive entropy  production does not change with time,  one can achieve Carnot efficiency at finite power. 
We may call this equality as steady entropy production condition if there is no heat leakage between the system and the reservoirs. Using the above equality, in the following section, we try to calculate the efficiency at maximum power from 
the minimally nonlinear irreversible heat engines under the non-tight coupling condition.

\section{Efficiency at maximum power under the non-tight coupling condition}
Under the non-tight coupling condition, $|q| \ne 1$ and hence the leakage term
becomes non-zero ($g \ne 0$) for $L_{ij}>0$.  Since the integration constant 
$A=\beta g$ as obtained from Eq.(\ref{Aval}) also depends  on $\tau$
and $\beta=X_2T_c$ is a function of $T$ alone, one can expect 
$g$ should   also depends on $\tau$  for a given $\beta$.
For the simplest choice, we take $g= B/(\beta \tau^2)$, where B is 
a constant and using this value of $g \ne 0$ in Eq.(\ref{j3min1}), we get 
\begin{eqnarray}\label{j3minnon}
J_3(T,\dot{T}) &=& \frac{B}{\tau^2}+ (X_2T_c -1) C_v \dot{T}. 
\end{eqnarray}
Integrating the above equation from $0$ to $\tau$ and using Eqs.(\ref{wbound}),(\ref{delS}) and (\ref{powval})  we obtain
$\int_0^{\tau} {J_3(T,\dot{T})} dt=\frac{B}{\tau^2} \int_{0}^{\tau} dt -T_c\int_{T_h}^{T_c} {\frac{C_v}{T} dT}$. 
\begin{eqnarray}\label{j3nons1}
\int_0^{\tau} {J_3(T,\dot{T})} dt =\frac{B}{\tau} +T_c (S_h-S_c).
\end{eqnarray}
and the total power
\begin{eqnarray*}\label{pownontmp}
P&=&\frac{1}{\tau} \bigg(U_h-U_c-\frac{B}{\tau} -T_c (S_h-S_c)\bigg). 
\end{eqnarray*}
\begin{eqnarray}\label{pownon}
P&=&\frac{1}{\tau} \bigg(E-\frac{B}{\tau}\bigg). 
\end{eqnarray}
In order to find out the value of $\tau=\tau^*$ in which the total power is maximum, one can   
maximize Eq.(\ref{pownon}) with respect to $\tau$ as
\begin{eqnarray}\label{pownonmx}
\frac{dP}{d\tau}&=& \frac{-E}{\tau^2} + \frac{2B}{\tau^3}=0 
\end{eqnarray}
and obtain
\begin{eqnarray}\label{taumax}
\tau^*&=& \frac{2B}{E}. 
\end{eqnarray}
With this value of $\tau^*$, we obtain the maximum power 
\begin{eqnarray}\label{powmax}
P^*&=&\frac{E^2}{4B}. 
\end{eqnarray}
Using Eqs.(\ref{etabound}) and (\ref{powmax}) , we obtain the work output and the efficiency at 
maximum power under the non-tight coupling condition as
\begin{eqnarray}\label{Wmax}
W^*=P^* \tau^*=\frac{E}{2}, 
\end{eqnarray}
\begin{eqnarray}\label{effmax}
\eta^*=\frac {W^*}{U_h - U_c} = \frac{1}{2} \eta_{max}.
\end{eqnarray}

This result shows that the efficiency at maximum power is equal to  half of the reversible efficiency
and the corresponding  work is half the exergy.
Our final result is exactly the same as the one obtained  earlier for the study of exergy \cite{bathfinite}
in the case of linear irreversible heat engines under the tight coupling condition.  
This shows that the efficiency and the work at maximum power obtained from the linear irreversible heat 
engines under the tight coupling \cite{bathfinite} is a special case of the 
efficiency at maximum power obtained from the minimally nonlinear irreversible 
heat engine under the non-tight coupling condition for a specific value of $g$.

\section{Conclusion}

Using the general formulation of the irreversible thermodynamics, we studied the 
optimized work and the efficiency of minimally nonlinear irreversible heat engines 
operating  between  finite sized hot and infinite sized cold reservoirs. 
We obtained the necessary condition to achieve an optimized work output. 
Our condition obtained in the case of minimally nonlinear irreversible model resembles with 
the one obtained recently  \cite{nonlinbath} for the generalized study of the 
irreversible heat engines in the nonlinear regime.

We used the optimization condition, Eq.(\ref{ELconst}), and calculated the maximum work and efficiency 
of the minimally nonlinear irreversible heat engines. Earlier studies for the irreversible heat engines 
showed that the tight coupling condition serves as an upper bound of the efficiency 
at maximum power. Interestingly, our result showed that the 
reversible efficiency can be achieved at finite power for nonlinear irreversible heat engine
under the tight coupling condition. Our results also showed that the reversible efficiency 
obtained from the nonlinear irreversible heat engines in the tight coupling condition
is not necessarily to be in the regime of maximum or zero power output.

We have also calculated the efficiency at maximum power from the nonlinear irreversible  heat engine 
under the non-tight coupling condition for a specific value of $g$ and found that 
the efficiency at maximum power is equal to the half the reversible efficiency
and the corresponding work is half the exergy. This result is exactly the
same as  the efficiency and the work at maximum power obtained from the linear irreversible 
heat engines under the tight coupling  condition \cite{bathfinite}.

Our result showed that the reversible efficiency at finite power is theoretically possible 
for heat engines working in the nonlinear regime. The validity of this result is based mainly on the assumption of the presence of symmetry in the Onsager coefficient and 
the rate of heat transfer from hot to cold bath obeys Fourier law.
Our future work will focus the alteration of the present analysis for the non-symmetric Onsager coefficient \cite{minthermoelect}
and the anomalous heat transfer of the systems in which the Fourier's law is in general not valid  \cite{nofourier}.

\section*{Appendix:}
$k(T) =\frac{\beta (T) - X_2(T) T_c}{a_1(T)}$ and 
$p(T,\dot{T})=1+2a_1(T) \big( g(T)+ C_v (T)\dot{T} \big)$.
The partial differentiation of $p$ 
with respect to $T$ and $\dot{T}$ is given by
\begin{eqnarray}\label{pdiff1}
\frac{\partial{p}}{\partial{\dot{T}}}&=&2a_1C_v 
\end{eqnarray}

\begin{eqnarray}\label{pdiff2}
\frac{\partial{p}}{\partial{T}}&=&2\frac{\partial{(a_1g)}}{\partial{T}}+2\dot{T}\frac{\partial{(a_1C_v)}}{\partial{T}}.
\end{eqnarray}

By using Eqs.(\ref{j3fsub2} - \ref{pdiff2}) one  can calculate
\begin{eqnarray}\label{j3T}
\frac{\partial{J_3}}{\partial{T}}&=&\frac{\partial{}}{\partial{T}} \bigg(k[ 1 + \sqrt{p} ] + \beta g \bigg)  \\ \nonumber
 &+& \dot{T}\frac{\partial}{\partial T }\bigg((\beta -1)C_v \bigg). 
\end{eqnarray}

\begin{eqnarray}\label{j3Tdot}
\frac{\partial{J_3}}{\partial{\dot{T}}}&=&  \frac{k}{\sqrt{p}}a_1C_v+(\beta -1)C_v.
\end{eqnarray}

\begin{eqnarray}\label{j3Tddot}
\frac{\partial}{\partial{\dot{T}}} \bigg(\frac{\partial{J_3}}{\partial{\dot{T}}} \bigg) &=&  -\frac{k}{p^{3/2}}a_1^2C_v^2.
\end{eqnarray}

\begin{eqnarray}\label{j3TTdot}
\frac{\partial}{\partial T} \bigg(\frac{\partial J_3}{\partial \dot{T}} \bigg) &=& 
 \frac{\partial}{\partial T} \bigg(\frac{k}{\sqrt{p}}a_1C_v \bigg) \\ \nonumber
&+&\frac{\partial}{\partial T} \bigg((\beta -1)C_v \bigg). 
\end{eqnarray}

For optimization Eq.(\ref{ELeq}) can be rewritten in terms of T(t) as  
\begin{eqnarray}\label{ELeq1}
\ddot{T} \frac{\partial}{\partial \dot{T}}\bigg(\frac{\partial{J_3}}{\partial \dot{T}}\bigg)
+\dot{T} \frac{\partial}{\partial T}\bigg(\frac{\partial{J_3}}{\partial \dot{T}}\bigg)-\frac{\partial{J_3}}{\partial{T}}=0.  
\end{eqnarray}
By using Eqs.(\ref{j3T} -\ref{j3TTdot}) in the above equation one can obtain 
\begin{eqnarray}\label{ELeq2}
 -\ddot{T}  \frac{ka_1^2C_v^2}{p^{3/2}} &+& 
 \dot{T}\frac{\partial}{\partial T}\bigg(\frac{k}{\sqrt{p}} a_1C_v \bigg)  \\ \nonumber
&-&\frac{\partial}{\partial T}\bigg(k[1 + \sqrt{p}] + \beta g \bigg)=0.  
\end{eqnarray}

Since
\begin{eqnarray}\label{Tdoteq}
\frac{d}{dt}\bigg( \frac{k}{\sqrt{p}} a_1C_v \bigg)= -\ddot{T}  \frac{ka_1^2C_v^2}{p^{3/2}}
 + \dot{T} \frac{\partial}{\partial T} \bigg(\frac{k}{\sqrt{p}} a_1C_v \bigg)   
\end{eqnarray}
and 
\begin{eqnarray}\label{Tdoteq1}
\frac{\partial}{\partial \dot{T}} \bigg(k[1 + \sqrt{p}] + \beta g \bigg) = \frac{k}{\sqrt{p}} a_1C_v,
\end{eqnarray}
Eq.(\ref{ELeq2}) can be rewritten as
\begin{eqnarray}\label{ELfin1_appendix}
\frac{d}{dt}\bigg( \frac{\partial Y} {\partial \dot{T}} \bigg) - \frac{\partial Y}{\partial T} =0, 
\end{eqnarray}
where $Y(T,\dot{T})=k[1 + \sqrt{p}]+\beta g$.
We also get the same type of  Eq.(\ref{ELfin1_appendix}) for the other value of $J_1=J_1^{-}$
with $Y(T,\dot{T})=k[1 - \sqrt{p}]+\beta g$.
Therefore for two different values of $J_1=J_1^{\pm}$, one can get the same equation.

\section*{Acknowledgments:}

I thank R. Arun for the critical reading of the manuscript. I also thank the anonymous referees
for their critical comments and valuable suggestions.

\end{document}